\documentstyle[aps]{revtex}
\title{Glassy behaviour of the ``Site Frustrated Percolation'' model}
\author{Silvia Scarpetta$^{a,b,c}$,
        Antonio de Candia$^{c,d}$, Antonio Coniglio$^{c,d}$}
\address{${}^a$Dipartimento di Fisica Teorica, via S. Allende,
         84081 Baronissi (SA)}
\address{${}^b$INFM, Sezione di Salerno}
\address{${}^c$Dipartimento di Fisica, Mostra d'Oltremare
         pad. 19, 80125 Napoli}
\address{${}^d$INFM, Sezione di Napoli}
\begin{document}
\maketitle
\begin{abstract}
The dynamical properties of the  ``Site Frustrated Percolation'' model
are investigated and compared with those of glass forming liquids.
 When the density of the particles on the lattice becomes
high enough, the dynamics of the model becomes very slow,
due to geometrical constraints,
and rearrangement on large scales is needed to allow relaxation.
The autocorrelation functions,
the specific volume for different cooling rates, and the mean square
displacement are evaluated, and are found to exhibit  glassy behaviour.
\end{abstract}
%
%%%%%%%%%%%%%%%%%%%%%%%%%%%%%%%%%%%%%%%%%%%%%%%%%%%%%%%%%%%%%%%%%%%%%
%
\section{Introduction}
Frustration plays a central role in many complex systems,
such as spin glasses\cite{binyoung} and glass forming liquids\cite{edangell}.
In spin glasses frustration arises because ferromagnetic and
antiferromagnetic
interactions  are distributed in such a way that  the spins cannot
simultaneously satisfy all interactions.
In glass forming liquids frustration arises when local arrangements of
molecules  kinetically prevent all the molecules from reaching ordered
close-packed configurations.
\par
In this paper we study a lattice-gas model which contains frustration
as an essential ingredient.
This model is the site version of the bond frustrated percolation model, which
was defined in the context of the generalization of the
Kasteleyn-Fortuin\cite{Kasteleyn}
Coniglio-Klein\cite{CKlein} cluster formulation to  the spin glass model
\cite{Coniglio1}.

\par
More precisely consider the Ising spin glass Hamiltonian
\begin{equation}
H=-J\sum_{\langle{ij}\rangle}\epsilon_{ij}S_iS_j,
\end{equation}
where $\epsilon_{ij}$ are quenched random interactions which assume
the values $\pm 1$ with equal probability.
By introducing  bond variables between each nearest-neighbour pair of spins,
it is possible to show that
the spin glass partition function can be expressed as a sum over
bond configurations on the lattice \cite{Coniglio1}
\begin{equation}
Z={\sum_C}^{\ast}e^{\beta\mu n(C)}q^{N(C)},
\label{spinglass_Z}
\end{equation}
where $q=2$ is the multiplicity of the spins,
$\beta\mu = \ln(e^{q\beta J}-1)$,
$n(C)$ and $N(C)$ are respectively the number of bonds and
the number of clusters in
the bond configuration $C$.
The asterisk
in (\ref{spinglass_Z}) means that the sum extends over all the bond
configurations
that do not contain a  ``frustrated loop''. This is defined as
a closed path of bonds
which contains an odd number of antiferromagnetic interactions.
In this formalism the spin glass problem is mapped on to a geometrical
problem, which for $T=0$ corresponds to  pack the highest number
of bonds without closing frustrated loops.
\par
For general values of $q$ the  model has been called
``q-Bond Frustrated Percolation ''
\cite{Coniglio2} and the partition function
(\ref{spinglass_Z}) can be obtained from an hamiltonian model
in which, besides the Ising spins interacting in a spin glass
way, there are Potts spins in each site, interacting ferromagnetically,
with multiplicity $s=q/2$ \cite{Cataudella}.

The model exhibits two critical points for each value of $q$:
one at  higher temperature $T_p(q)$,
corresponding to the percolation transition, in the same universality
class of the ferromagnetic Potts model with multiplicity  $s=q/2$,
and one at lower temperature
$T_{\rm SG}(q)$, in the same universality class of the Ising spin glass
transition.
This has been
verified by renormalization methods \cite{CPezzella},
and numerically \cite{AdC}.

It has also been shown \cite{Cataudella,CPezzella},
that for $q\neq 2$ the percolation
transition corresponds to a singularity in the partition function,
and thus it corresponds to  a real thermodynamic transition.

\par
Each critical point is characterized by a diverging length, associated
with the quantities\cite{Cataudella,Coniglio2}:

\begin{equation}
p_{ij} = p_{ij}^{+} + p_{ij}^{-}
\label{pij}
\end{equation}

and

\begin{equation}
g_{ij} = p_{ij}^{+} - p_{ij}^{-}
\label{gij}
\end{equation}

Here $p_{ij}^{+}$ $(p_{ij}^{-})$ is the probability that 1) sites
$i$ and $j$ are connected by at least one path of bonds, and 2) the
phase $\eta_{ij}$ defined as the
product over all the signs $\epsilon _{mn}$ along the path
connecting $i$ and $j$ is $+1$ $(-1)$\cite{foot}.

The length $\xi_p$ associated with the pair connectedness function
$\overline{p_{ij}}$ diverges at the percolation critical point,
while the length $\xi$ associated with $\overline{g_{ij}^2}$ diverges
at $T_{\rm SG}(q)$
(the bar represents the
average over all possible interaction configurations $\{\epsilon_{ij}\}$).
For $q=2$  the quantity $g_{ij}$  in Eq. \ref{gij} coincides with
the spin spin pair correlation function.

However we can give a geometrical interpretation of the second length $\xi$.
The second transition, like the quantum percolation
transition\cite{aa}, occurs at
a bond density higher than the usual percolation transition, due to the
 interference of paths with different phases.
At high density the number of allowed configurations is extremely reduced
and most of the configurations  connecting  a given pair of sites $i$ and $j$
will have a common path. We call this path a ``quasi frozen'' path,
since in a dynamical sequence of configurations exploring the allowed
phase space, this path will be present most of the time.
In conclusion the major
contribution to $g_{ij}$ is due to those configurations in which $i$
and $j$ are connected by quasifrozen bonds\cite{Coniglio2,jan,sg}
and $|\overline{g_{ij}}|$ roughly coincides with the probability that $i$
and $j$ belong to the same quasifrozen cluster, and the length $\xi$
associated with $\overline{g_{ij}^2}$ roughly represents the linear
dimension of these clusters.

For $q=1$ (\ref{spinglass_Z})  reduces to the bond frustrated percolation
model, in which
bonds are randomly distributed on the lattice, with the only
constraint that configurations of bonds which contains at least one
frustrated loop are not allowed\cite{ap}.

Thus the model incorporates in the simplest way the concepts of geometrical
frustration and could be applied to those systems, such as glass forming
liquids, where geometrical frustration due to
packing problems plays a central role.
We note that in this model the disorder is quenched
while in glass   forming liquids this is not the case. However
at low temperature or high density the relaxation times are so large
that the disorder may be considered frozen in.
\par
To better describe the particle nature of glassy systems, we have
introduced in this paper the
site version of the frustrated percolation model, in which
sites can be full or empty. Full sites can be seen as ``particles'',
and empty sites as vacancies.
We leave unchanged
the underlying structure of the interactions, and the constraint that no
frustrated loop can be fully occupied.
We consider in this paper only the $q=1$ case.
\par
In sect. \ref{sec_model} we introduce the model, and in sect.
\ref{sec_montecarlo} we describe
two types of Monte Carlo dynamics (thermalization and diffusion
dynamics), used to simulate the model.
We have studied the static properties of the system in
sect. \ref{sec_percolation}. In particular we have verified that
the percolation transition is in the same universality class
of the $s=1/2$ ferromagnetic Potts model.
The cooling rate dependence of the specific volume is considered in
sect. \ref{sec_cooling}, while in sect. \ref{sec_diffusion} we
analyse the diffusion dynamics of the model, evaluating the mean
square displacement of the particles and the diffusion coefficient.
Finally, in sect. \ref{sec_relaxation} we evaluate the autocorrelation
functions of density fluctuations, both in thermalization and
diffusion dynamics, which show a behaviour typical
of glass forming liquids. In the appendix we describe in more detail
the Monte Carlo procedure.
\section{The ``Site Frustrated Percolation'' model}
\label{sec_model}
In the site frustrated percolation (SFP) model particles are introduced on the
vertices of
a regular lattice. We assign to each edge of
the lattice an interaction $\epsilon_{ij}=\pm 1$, positive or negative
according to a random quenched distribution.
Like in spin glasses a frustrated loop is a closed path of edges
which contains an odd number of negative
interactions.
A particle configuration is forbidden only
if it contains a fully occupied frustrated loop\cite{nic} (see Fig.
\ref{site_percolation}).
Thus we assign zero weight to all configurations of particles which
contain at least one fully occupied frustrated loop, and
to all other configurations a weight
\begin{equation}
W(C)=e^{\beta\mu n(C)},
\end{equation}
where $n(C)$ is the number of particles in the configuration $C$,
$\mu$ is the chemical potential of the particles and $\beta=1/k_BT$.
\par
The partition function of the model is given by
\begin{equation}
Z = {\sum_C}^\ast e^{\beta\mu n(C)},
\label{site_Z}
\end{equation}
where the sum excludes forbidden configurations.
\par
Note that there is only one independent parameter $\beta\mu$, which can vary
from $-\infty$ to $\infty$.
Frustration prevents the system from reaching maximum density $\rho=1$
for $\beta\mu\to\infty$: when $\beta\mu$ varies from $-\infty$ to $\infty$,
the density $\rho=\langle{n}\rangle$ varies from $0$ to $\rho_{\rm max}<1$.
\par
The SFP model is expected to have two critical
points, in analogy to the bond case \cite{Cataudella,CPezzella,AdC}.
Fixing the value of $\mu$ to a positive value, and varying the temperature
from high to low values, there is a first critical point at a temperature
$T_p$, corresponding to the percolation transition, and one at lower
temperature $T_{\rm SG}$, corresponding to a spin glass transition.
The percolation transition is expected to be in the same universality class
of the ferromagnetic $s=1/2$ ferromagnetic Potts model, while
in two
dimensions  $T_{\rm SG}$ is expected to occur at $T=0$.
\par
At low temperature the model is expected to have the
characteristic features of a spin glass system, such as a rough free energy
landscape, very long relaxation times due to the high free energy barriers,
and many ground states at $T=0$.
\section{Monte Carlo dynamics}
\label{sec_montecarlo}
We have realised two types of Monte Carlo dynamics to simulate the SFP model.
The first, which we call ``thermalization dynamics'', proceeds through the
following steps:
\begin{itemize}
\item pick up a site at random;
\item if the site is filled by a particle, destroy that particle with
probability $P_{-}$, or leave the site filled with probability
$(1-P_{-})$;
\item if the site is empty, leave it empty
if a new particle placed in that site would
close a frustrated loop otherwise,
if the particle would not close a frustrated
loop,
create a new particle in that
site with probability $P_{+}$, or leave the site empty with
probability $(1-P_{+})$.
\end{itemize}
Looking at the partition function (\ref{site_Z}), it is easy to
verify that the following probabilities of creating or destroying a particle
satisfy the principle of detailed balance:
\begin{mathletters}
\begin{equation}
\left\{
\begin{array}{rcl}
P_{-}&=&1\\
P_{+}&=&e^{\beta\mu}
\end{array}
\right.
\hspace{3cm}\mbox{for $\beta\mu<0$;}
\end{equation}
\begin{equation}
\left\{
\begin{array}{rcl}
P_{-}&=&e^{-\beta\mu}\\
P_{+}&=&1
\end{array}
\right.
\hspace{3cm}\mbox{for $\beta\mu>0$.}
\end{equation}
\end{mathletters}
This dynamics is clearly ergodic, because we can go from a permitted
configuration $A$ to a permitted configuration $B$, first destroying
the particles belonging to $A$, and then creating those
belonging to $B$.
\par
The difficult step here is to verify that a new particle doesn't close a
frustrated loop, since this involves a non local check.
The procedure is described in the appendix.
\par

The second type of Monte Carlo dynamics, which we call ``diffusion dynamics'',
starts from a site configuration with some density $\rho$ of particles,
obtained by thermalization dynamics, and proceeds
by letting the particles diffuse conserving their number.
At each step a site is chosen at random; if the site is filled,
we make an attempt to move the particle to a NN site. The move is
accepted if the probed site is empty and no frustrated loop is closed.
\par
\section{Percolation transition and equilibrium density}
\label{sec_percolation}
In this section we analyse the percolation properties of the model.
The analysis of the data confirms that the percolation
transition, for the site frustrated percolation problem, is in the same
universality class of the ferromagnetic $s=1/2$ Potts model,
as was formerly verified
for the bond problem \cite{AdC}.
\par
We have used the histogram method for analyzing data \cite{Swendsen}.
For various lattice sizes, we simulated the system using thermalization
dynamics for 10 temperatures around the percolation point. For each
temperature we reached equilibrium by taking $10^4$ steps, and then
evaluated the histograms of the following quantities, taking $10^5$ steps:
\begin{itemize}
\item density of particles $\rho$;
\item probability of existence of a spanning cluster $P_\infty$;
\item mean cluster size $\chi$.
\end{itemize}
The mean cluster size is defined as
\begin{equation}
\chi = \frac{1}{N}\sum_s s^2n_s,
\end{equation}
where $n_s$ is the number of clusters having size $s$ in the system.
\par
Using the histogram method \cite{Swendsen}, we evaluated the
values of these three quantities for an entire interval of the parameter
$\beta\mu$.
In Fig. \ref{percolation_FSS} is shown the finite size scaling of
$P_\infty$ and of $\chi$, from which it is possible to extract
the values of the critical temperature, and of the critical
exponents $\nu$ and $\gamma$ \cite{Binder}.
\par
The data are perfectly compatible with the values $\nu^{-1}=0.56$,
$\gamma=3.27$, of the $s=1/2$ Potts model \cite{Wu}.
The critical (percolation) value for $\beta\mu$ is found to be
$(\beta\mu)_p=1.2$, which corresponds to a density
$\rho_p\simeq 0.6$ which is, within the error, not different
from the percolation density of the standard random site percolation,
$\rho_p=0.593$ \cite{Stauffer}.
\par
We evaluated the equilibrium density of the SFP, varying  the parameter
$\beta\mu$ from $\beta\mu=-15$ to  $\beta\mu=15$,
with rate $3\times 10^{-5}$
step$^{-1}$. The result is shown in Fig. \ref{eq_density}.
For comparison we plot also the
equilibrium density of the unfrustrated model, which is given by
$\rho=1/(1+e^{-\beta\mu})$.
This curve fits quite well the low density part, since for low density
the effect of frustration is negligible.
The high density part instead is better fitted
by the function
\begin{equation}
\rho=\frac{\rho_{\rm max}}{1+e^{-a-b(\beta\mu)}}
\label{density_fit}
\end{equation}
where $\rho_{\rm max}=0.788$, $a=0.323$, $b=0.719$.
Note from
Fig. \ref{eq_density}
that the crossover occurs at a value of
$\beta\mu$, close to the percolation threshold.

\section{Cooling rate dependence of the density}
\label{sec_cooling}
The effect of frustration prevents the system from {\em easily} reaching the
equilibrium density, especially at very high values of the parameter
$\beta\mu$, that is for $\beta\mu\rightarrow\infty$.
\par
Fixing $\mu$ to a positive value, and cooling the temperature from high to
low values with a finite cooling rate,
there is a temperature
$T_G$ at which the system goes out of equilibrium. Lowering further the
temperature, the density remains constant to the value corresponding to the
temperature $T_G$.
\par
Fig. \ref{cooling_rates} shows the temperature dependence
of the density for various cooling rates.
Temperature goes from $k_BT/\mu=0.5$ to $k_BT/\mu=0$, and
cooling rates range from $k_B\dot{T}/\mu=10^{-2}$ step$^{-1}$ to
$10^{-7}$ step$^{-1}$. One step is one update per site.
\par
This behaviour is experimentally observed in supercooled glass-forming
liquids, at the calorimetric glass transition temperature $T_G$,
when structural relaxation times become greater than the experimental
observation times \cite{Zarzycki}. As in glass forming liquids, in the model
we observe that the faster the cooling rate, the greater are the glass
transition temperature and the specific volume.
\section{The diffusion dynamics}
\label{sec_diffusion}
We have then studied the diffusion dynamics (conserved number of particles)
of the system.
The diffusion process is severely  hindered by kinetic
constraints at high density, and particles can diffuse through the
 system only by a large scale
cooperative rearrangement of many particles.
 At low density we expect that the effect of frustration on the diffusion
 is very weak because of abundance
of holes, and particles diffuse freely on the lattice.\par
We have evaluated the mean square displacement (MSD)
$\langle\Delta r(t)^2\rangle$ as a function of time $t$,
as shown in Fig. \ref{mean_sd}.
Each curve is obtained  averaging over all the particles
and over a time interval of 5000 steps.
As expected for low density the curves show a
linear behaviour, which corresponds to normal
diffusion.
For high densities, the MSD reaches a plateau
and then becomes again linearly  dependent on
time. This behaviour is typical of glass forming liquids and is also
observed in molecular dynamics simulations\cite{KobAndersen,sciortino}.
The crossover from the normal behaviour to the anomalous diffusion in our
model again
occurs at a density value close to percolation.
\par
We have extrapolated the diffusion coefficient values
$D=\langle\Delta r(t)^2\rangle /t$
from the long time
regime which are shown in Fig. \ref{diffusion_fit}.
The extrapolated values are very well fitted by a power law as
function of density.
\begin{equation}
D\sim (\rho -\rho_c)^{\gamma}
\label{D_fit}
\end{equation}
with $\rho_c \simeq 0.7874$ and ${\gamma}\simeq 1.5$.
\par
Note that $\rho_c$ is equal within the errors to $\rho_{\rm max}$,
the density corresponding to the ground state, as extracted from the fit
(\ref{density_fit}).
Thus the dynamic singularity
coincides in this model to the spin glass transition, which in two dimensions
occurs at $T=0$.
If we set $\rho_c=\rho_{\rm max}$, using Eq. (\ref{density_fit}) we obtain
from Eq. (\ref{D_fit})
\begin{equation}
D\sim \left(1-\frac{1}{1+e^{-a-b(\beta\mu)}}\right)^{\gamma}\sim
e^{-b\gamma(\beta\mu)}
\end{equation}
which shows that the diffusion coefficient goes to zero with an Arrhenius law
for $T\to 0$.
\par
The picture which emerges from the model is that particles
are trapped  most of the time in cages formed by their NN, and can diffuse
only into localized pockets. Only after a very long interval of time they have
the opportunity of escaping from the cage, falling in a ``next neighbour''
cage. This process forms a kind of random walk, reflected in the long time
linear dependence
of the MSD functions.
\par
This picture is enforced by direct observation of
the particles moving on the lattice. We have taken ``snapshots'' of the system
at particular times, for diffusion dynamics at two high values of the
density. Some particles leave tracks on the lattice, so we can analyze also
the path they have walked.
In Fig. \ref{diffusion_path} is shown the snapshot
 at $\rho =0.783$ after 2000 steps (a), after 4000 steps (b)and
at $\rho =0.785$ after
2000 steps (c)
\par
We mark with dots the particles that moved at least one time
(non frozen domains) during the time of the simulation; the filled circles
represents frozen domains.
Clusters of frozen particles prevent other particles from moving freely.
We note that frozen particles are clustered and we can distinguish
liquid-like zones (holes and mobile particles) from
solid-like zones (frozen clusters).
Frozen clusters are eroded as the time of the simulation gets longer,
while liquid-like zones grow.
\par
At the very high density, frozen particles
remain blocked for extremely long times.
The frozen cluster in Fig. \ref{diffusion_path}(c),
corresponding to density $\rho=0.785$,
is still frozen after $2\times 10^5$ steps which corresponds to the maximum
time we have observed.
We have checked that the large frozen cluster does not percolate in either
direction, even if the density of
the frozen particles is  much greater than the percolation
density of the random site percolation. We expect that the frozen particles
in fact percolate at
$T=0$ ( $\rho= \rho_{max}$)
since the linear dimension of the frozen cluster should correspond
roughly to the correlation length associated to $g_{ij}$ (\ref{gij}),
which for $d=2$ is expected to diverge at $T_{SG}=0$, as explained
in the introduction.
\section{The relaxation functions}
\label{sec_relaxation}
An important property that characterizes glassy behaviour is the form
of the relaxation functions \cite{Cummins}. We
have evaluated the relaxation functions of the system,
in the thermalization and in the diffusion dynamics.
For each temperature, we reach the equilibrium, and then evaluate
the relaxation functions
averaging on a time interval of $10^3-10^6$ steps.
\par
Fig. \ref{correl_term}  shows the density-density autocorrelation function
in the  thermalization dynamics, defined as
\begin{equation}
 F(t)=\frac{\langle\delta\rho (t)\delta\rho (0)\rangle}
{\langle\delta\rho^2\rangle},
\end{equation}
 where $\delta\rho(t)=\rho(t)-\langle\rho\rangle$.
\par
For diffusion dynamics,
we have studied the autocorrelation function of the density
fluctuations
\begin{equation}
F_{\bbox{k}}(t)=\frac{\langle\delta\rho_{\bbox{k}}(t)
\delta\rho_{-\bbox{k}}(0)\rangle}
{\langle\delta\rho_{\bbox{k}}\delta\rho_{-\bbox{k}}\rangle},
\end{equation}
where
\begin{equation}
\rho_{\bbox{k}}(t)=\sum_{\alpha} e^{i{\bbox{k}}\cdot{\bbox{r}}_\alpha(t)}
\end{equation}
 and  ${\bbox{r}}_\alpha$ is the position of the $\alpha$-th particle
in units of lattice constant.
The wave vector can take the values
${\bbox{k}}=\frac{2\pi}{L}{\bbox{n}}$, where $\bbox{n}$ has
integer components $n_x$ and $n_y$ ranging in $0\ldots L/2$.
The autocorrelation functions corresponding to $n_x=L/2$ and $n_y=0$
are reported in Fig. \ref{correl_diff}.
\par
Fig. \ref{correl_self} shows the self-part of the relaxation function
in the diffusion dynamics, defined as
\begin{equation}
F^s_{\bbox{k}}(t)= \frac{1}{n}
\langle\sum_{\alpha} e^{i{\bbox{k}}\cdot({\bbox{r}}_\alpha(t)-
{\bbox{r}}_\alpha(0))}\rangle,
\end{equation}
where $n$ is the number of particles.
\par
At low temperatures, below the percolation threshold, we observe the onset of
non-exponential decay. This is usually the sign of non-stochastic
cooperative relaxation in the system.
We emphasize that, for our model, the percolation transition corresponds
to a real thermodynamic transition, and therefore it is possible to expect,
below this point, a change in the dynamical properties of the system.
\par
At very high density it is evident the existence of different time regimes
as predicted by the {\em mode mode coupling} theory \cite{MCT_1,MCT_2}
of supercooled liquids
and observed both in some molecular dynamic simulations and in some
experimental measurement \cite{Cummins} on glass forming liquids.
There is a first short time relaxation, corresponding to relaxation inside
non frozen domains surrounded by a frozen ``cage'', and a long time
regime ($\alpha$ relaxation), corresponding to structural
rearrangement and final decay to equilibrium.
\section{Conclusions}
We have studied a frustrated lattice gas model
which it has been called site frustrated percolation since percolation
plays an important role. This model,despite its simplicity, shows a
complex dynamical behaviour.
Disorder and frustration  are its basic ingredients.
The  constraint of frustration prevents the system from
easily reaching high density, and inhibits the motion of particles
at high density.
\par
We observed that the dependence of the volume
as function of the temperature varies strongly with the cooling rate,
qualitatively in the same way as observed in real glass forming liquids.
The motion of particles in diffusion
dynamics is severely hindered by the frustration constraint, and
the relaxation process can take place only through large
scale rearrangement.
The cooperative nature of the motion is reflected in the behaviour
of the mean square displacement and
in the relaxation functions of the system.
The plateau observed in the self part of relaxation functions
(Fig. \ref{correl_self}) and in the mean square displacement
(Fig. \ref{mean_sd}), are intrinsically connected, and are the sign of
the glassy nature of the dynamical properties of the model.
In all these phenomena the crossover from a normal behaviour to an
anomalous behaviour occurs close to or at the percolation threshold.
\par
Since SFP seems to well describe the glass transition in glass forming
liquids, the model suggests that the presence of a percolation type transition
may be a general feature below which frustration effects start to be
manifested. This transition may be responsible of various precursor phenomena
\cite{sg},
such as onset of stretched exponentials, breakdown of Stokes-Einstein
relation, presence of spatial heterogeneity.
\par
The presence of a percolation transition well above the glass transition
has recently been discovered by Tomida and Egami \cite{Tomida} in a molecular
dynamic simulation of monatomic liquids. It is also interesting to note that
Kivelson et al. \cite{Kivelson} showed that the viscosity of 15 glass forming
liquids could be collapsed on one single curve, by assuming only one
characteristic temperature well above the glass transition.

Finally the ideal glass transition temperature,
characterized by the divergence
of the inverse of the diffusion coefficient, numerically seems consistent
with the divergence of the static length $\xi$ associated
to $g_{ij}$ (\ref{gij}).

Although we have discussed the site frustrated percolation model in
the context of glassy systems, the model
is rather generic and may be applied to other systems,
where geometrical frustration plays an essential role. In fact recently
the model has been successfully applied also to granular materials
\cite{hh}.

We would like to thank V. Cataudella, F. di Liberto, S.C.Glotzer,
M. Nicodemi and
U. Pezzella for interesting discussions. This work has been supported
in part by CNR.

\section*{Appendix}
As mentioned in sect. \ref{sec_montecarlo},
we describe here the procedure to check
whether a new particle, added to the system in a given configuration,
closes or not  a frustrated loop.
\par
Every nearest neighbour (NN) particle (occupied site) of the empty
site, that has been probed to be filled, belongs to a
(not necessarily distinct) cluster of connected sites.
If two particles NN to the empty site
belong to the same connected cluster, then a new particle
filling that site closes a loop.
More precisely, if $z_n$ is the number
of particles NN to the empty site, and $z_c$ is the number of
{\em distinct} clusters to which
they belong, than the total number of new loops closed
by the new particle is $\lambda=z_n - z_c$.
\par
The algorithm proceeds as follows:
\begin{itemize}
\item
we count the number $z_n$ of particles NN to the empty site which we want to
fill, and mark each of them as the root of a distinct cluster,
so at the beginning $z_c=z_n$, and $\lambda=0$;
\item
we grow every cluster in parallel, adding to it in recursive way its
still not visited NN particles, and marking them as belonging to that cluster;
\item if no NN particle to a cluster is found,
that cluster is marked as ``burnt'',
and is not considered anymore;
\item  if two clusters collide, we say we have found a loop, and the two
clusters merge to form a single cluster; then $\lambda$ is
increased by one, and $z_c$ is decreased by one.
\end{itemize}
Every new visited site is marked in two ways: with the label of the cluster to
which it belongs, and with the ``parity'', that is the number of
antiferromagnetic interactions walked through starting from the initial
empty site. We stop the iteration when one of the following circumstances
happens:
\begin{itemize}
\item two clusters collide and the parities don't correspond (that is one
is odd and the other is even): in this case we have found a frustrated loop;
\item the number of burnt clusters equals $z_c-1$ (only one cluster
is non-burnt):
in this case no other loop can be found.
\end{itemize}
When the density of particles is low, this algorithm is very fast, since the
clusters are very low sized, and are burnt within few iterations.
On the other side, when the density is high, the maximum number of
independent loop is found quickly as well. The algorithm can become
slow when the density of particles is intermediate, notably near the
percolation transition, because clusters in this case are very ramified.
In Fig. \ref{CPU_times} the CPU
time needed to do a single site update is shown for a square bidimensional
system, and for various lattice
dimensions. The maximum time is reached near the percolation transition, and
scales as
\begin{equation}
\tau_{\rm max} \propto N^{0.4\pm 0.05},
\end{equation}
where $N=L^2$ is the total number of sites.

%
%%%%%%%%%%%%%%%%   REFERENCES  %%%%%%%%%%%%%%%%%%%%%%%%%%%%%%%%%%%%%%%%
%

%
%%%%%%%%%%%%%%%%%%%%%%%%%%%  FIGURES  %%%%%%%%%%%%%%%%%%%%%%%%%%%%%%
%
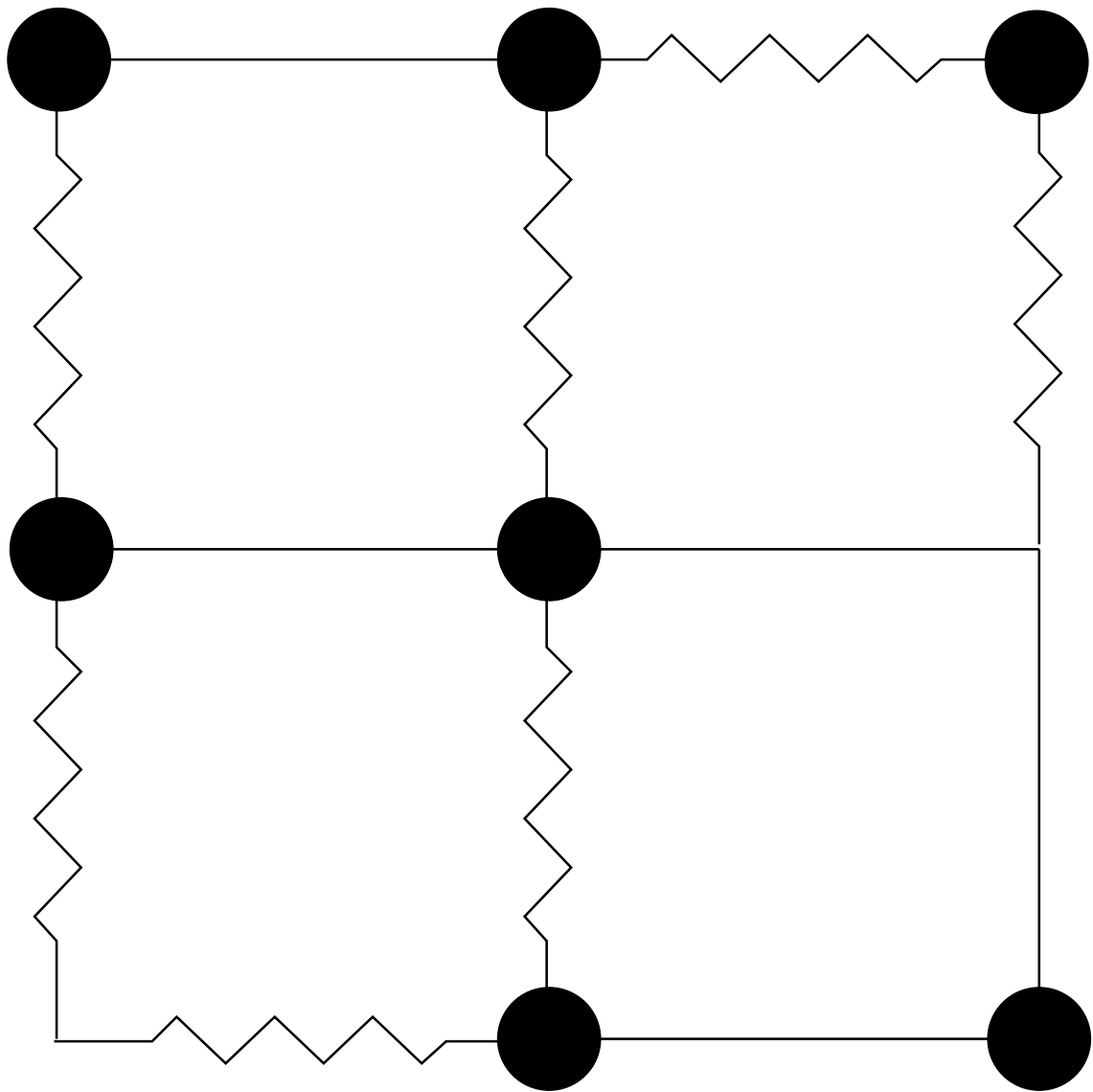
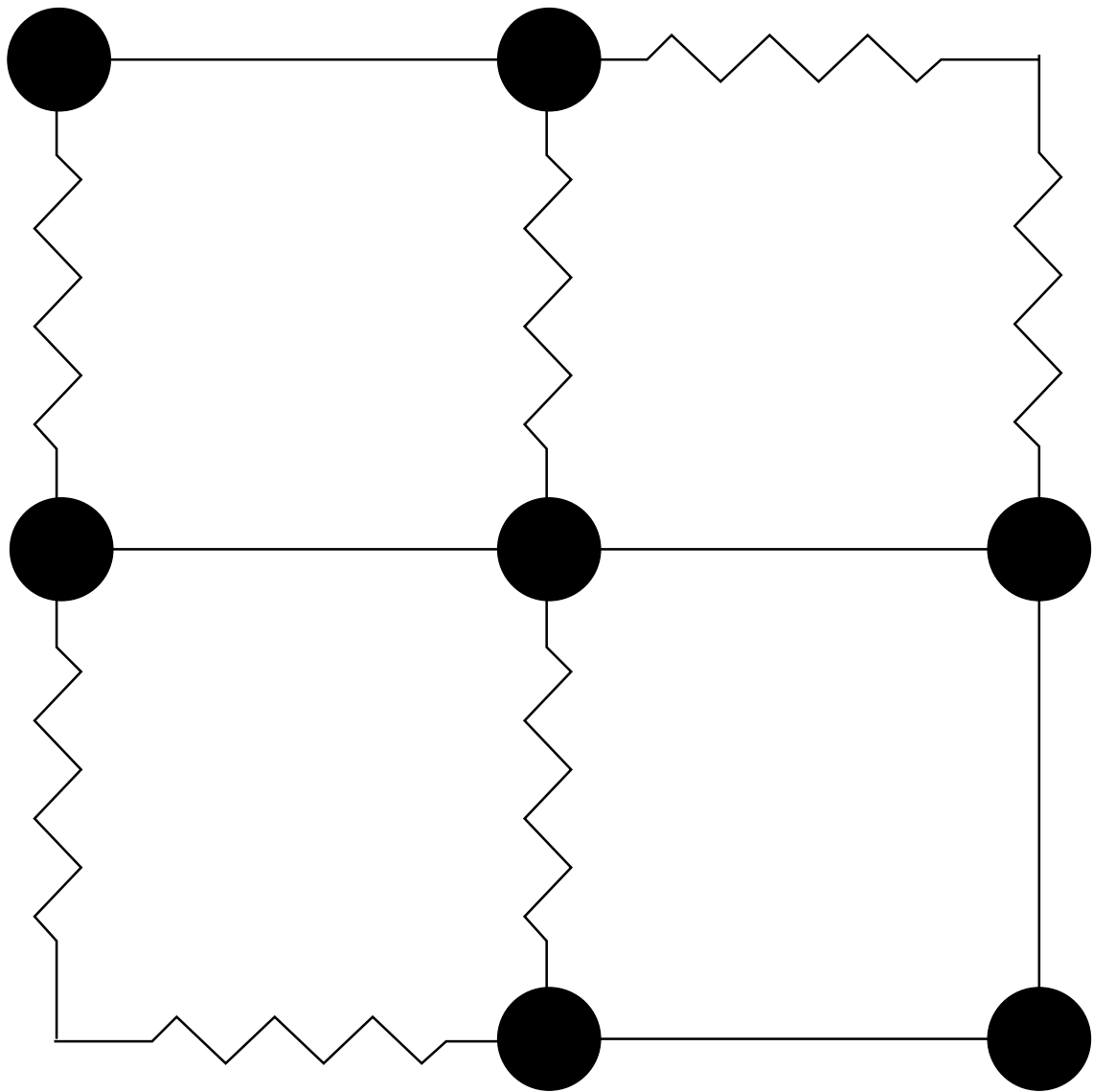
\begin{figure}
\caption{(a) Site configuration that does not contain frustrated loops
(permitted); (b) site configuration that contains a frustrated loop
(forbidden).}
\label{site_percolation}
\end{figure}
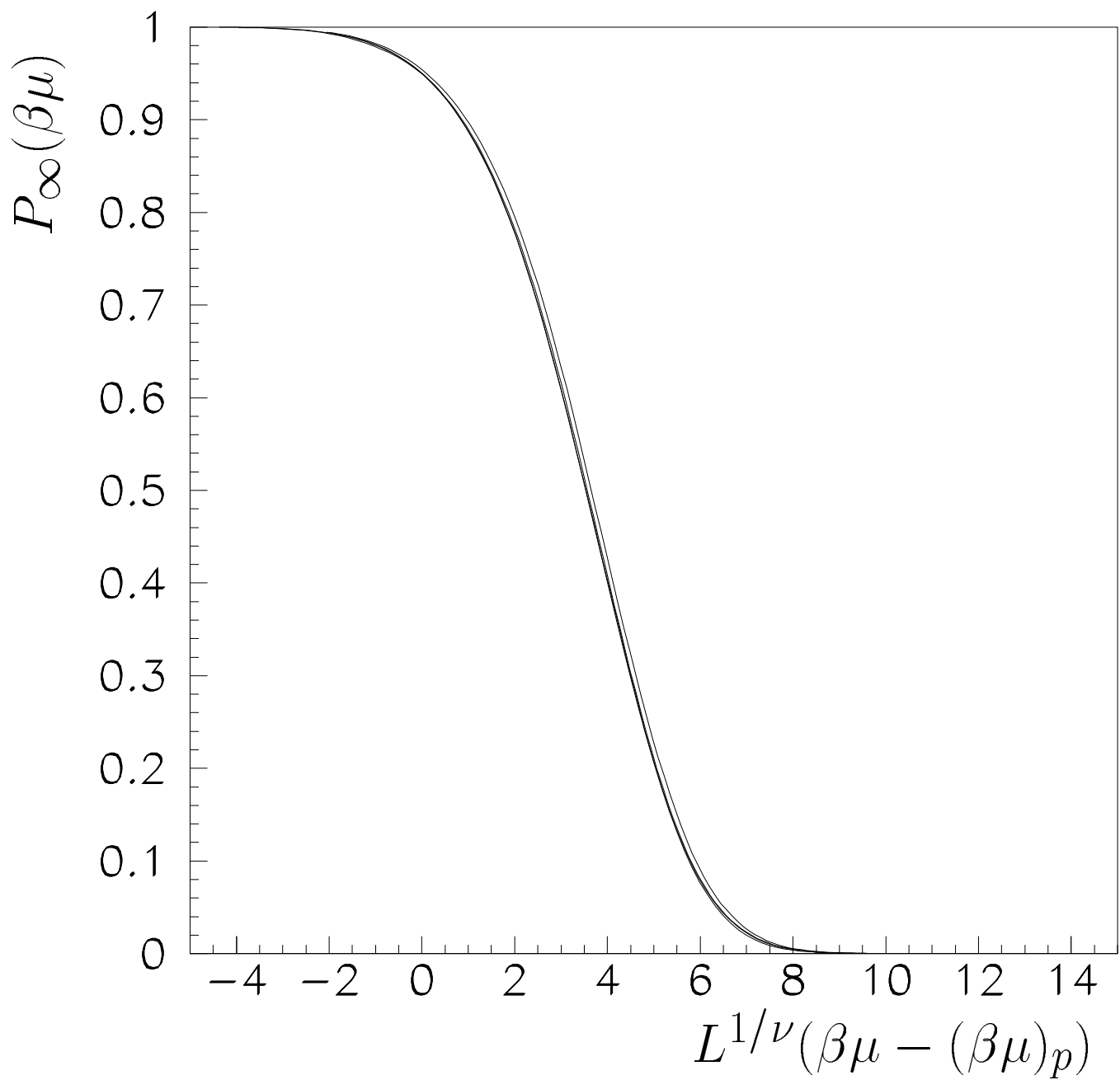
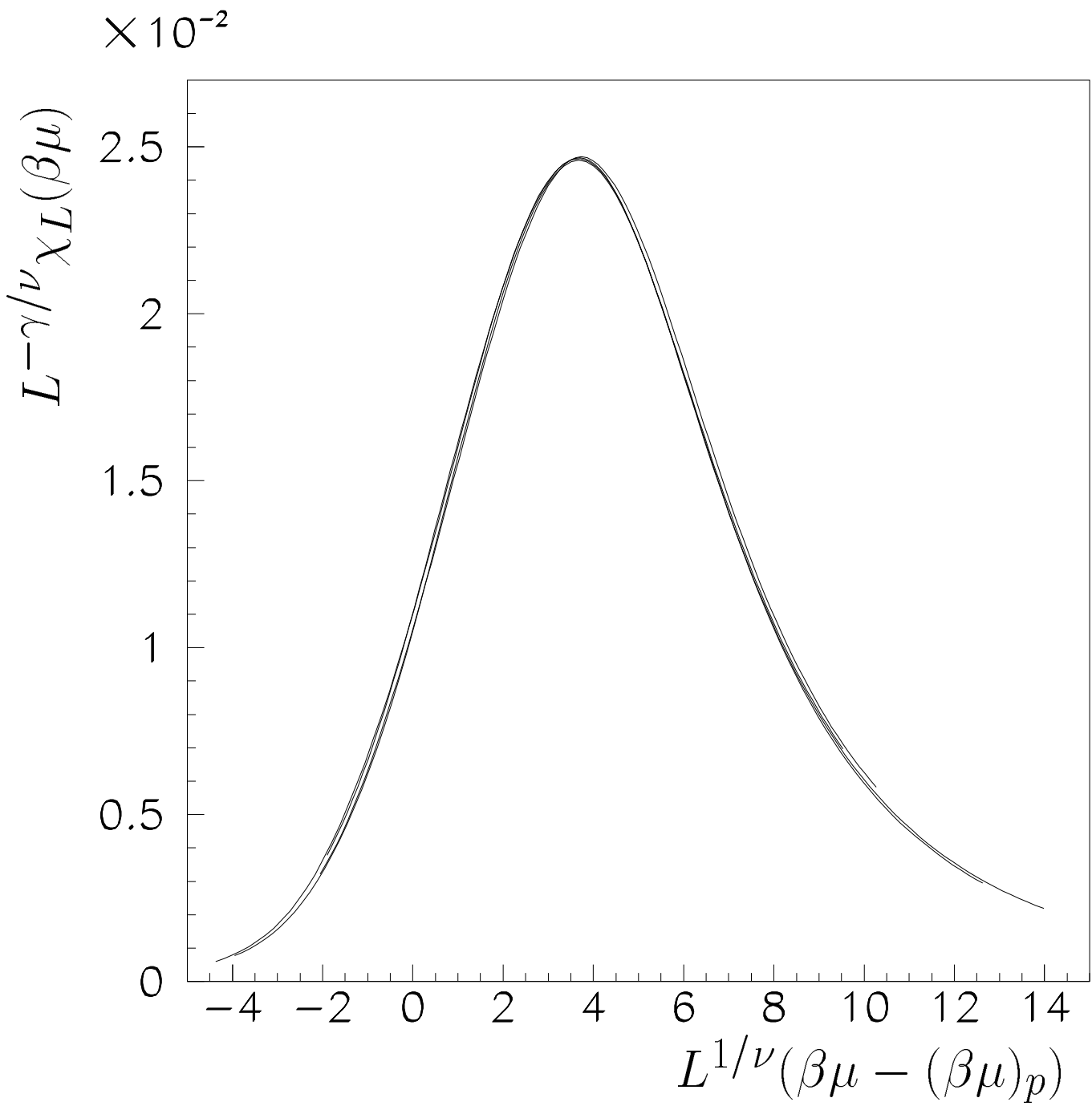
\begin{figure}
\caption{(a) Finite size scaling of the probability that a spanning cluster
exists; (b) finite size scaling of the mean cluster size.}
\label{percolation_FSS}
\end{figure}
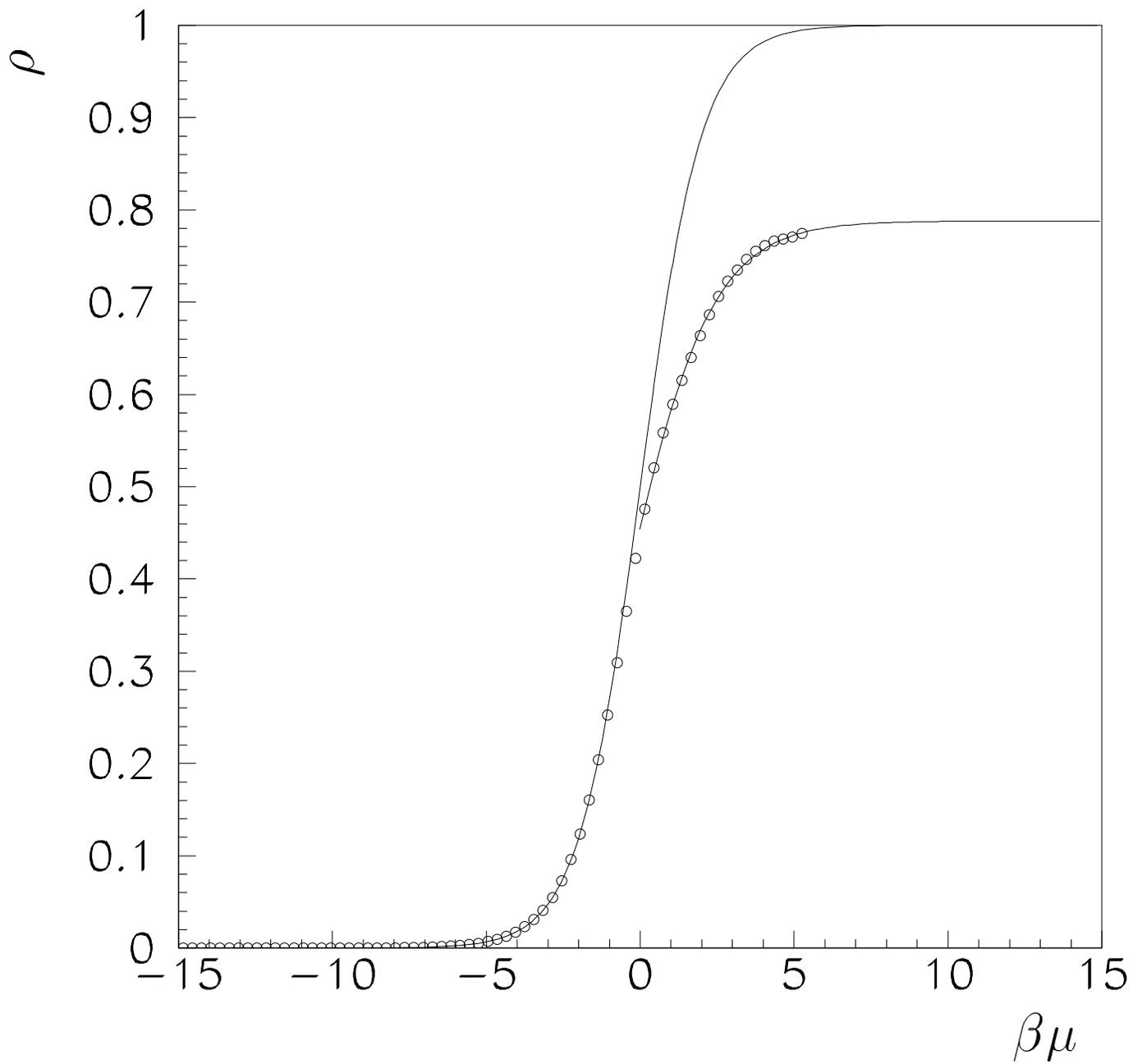
\begin{figure}
\caption{The equilibrium density of the SFP as function of
$\beta\mu$ (circles), together with the fit function (lower curve).
Upper curve is the density of the unfrustrated model.}
\label{eq_density}
\end{figure}
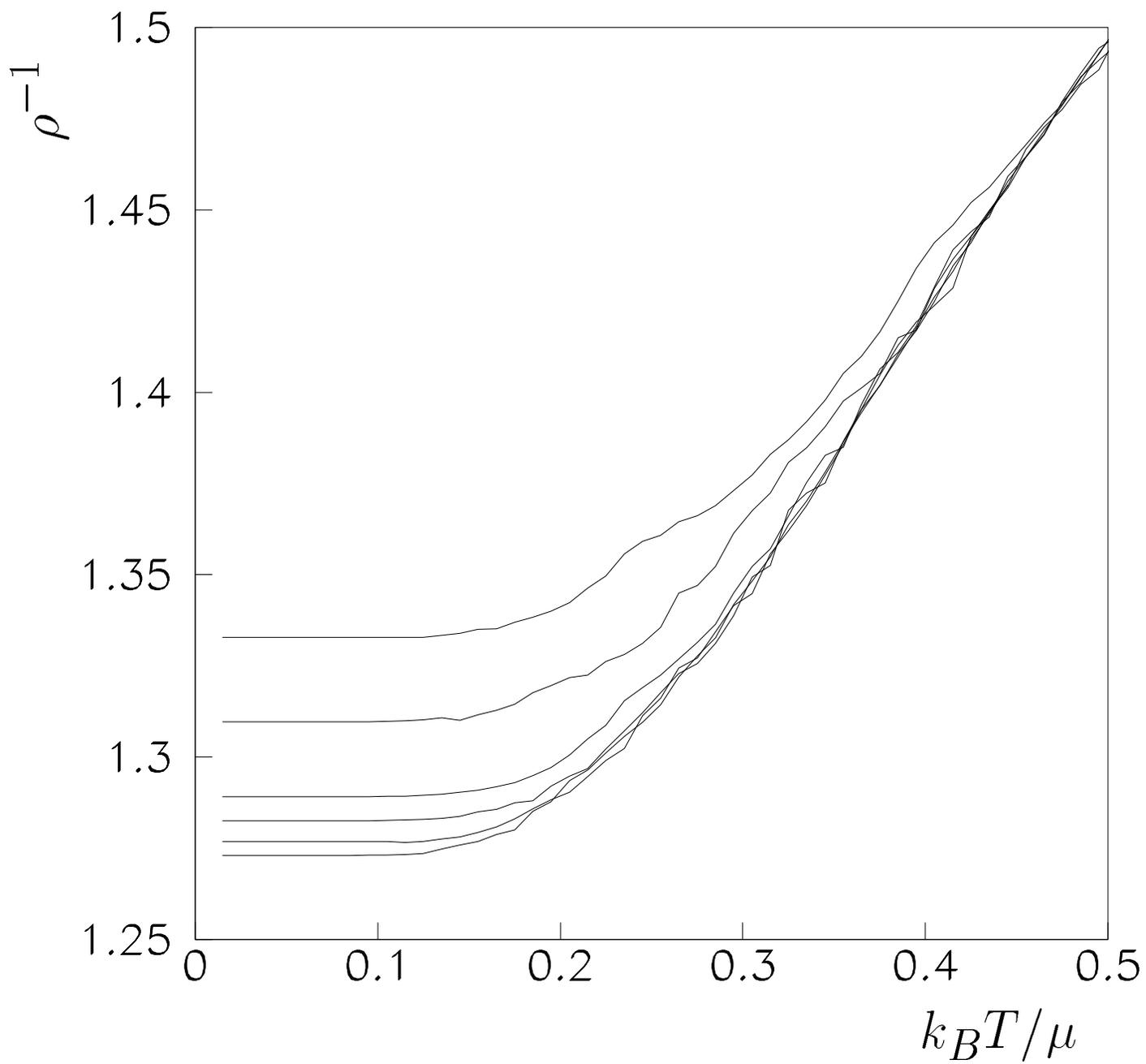
\begin{figure}
\caption{Temperature dependence of the specific volume in the SFP model
($32\times 32$ square lattice), for various cooling rates. From
upper curves to lower ones:
$k_B\dot{T}/\mu=10^{-2}$, $10^{-3}$, $10^{-4}$, $10^{-5}$, $10^{-6}$,
$10^{-7}$ step$^{-1}$.}
\label{cooling_rates}
\end{figure}
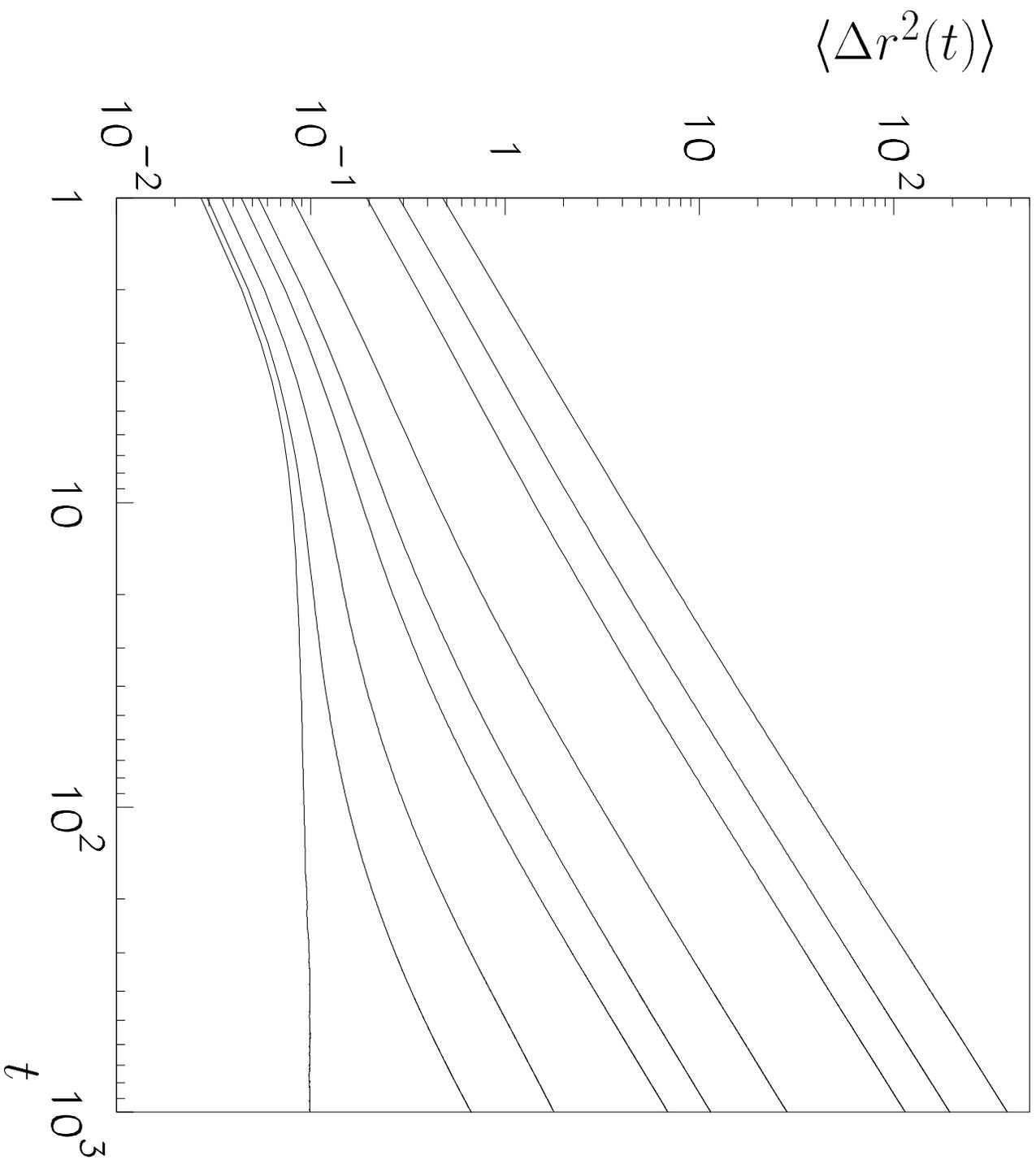
\begin{figure}
\caption{Mean square displacement in the SFP model
 on a square lattice with $L=32$,
for densities (from upper curves to lower ones): $\rho=0.452$, $0.586$,
$0.730$, $0.756$, $0.766$, $0.777$, $0.784$, $0.785$.}
\label{mean_sd}
\end{figure}
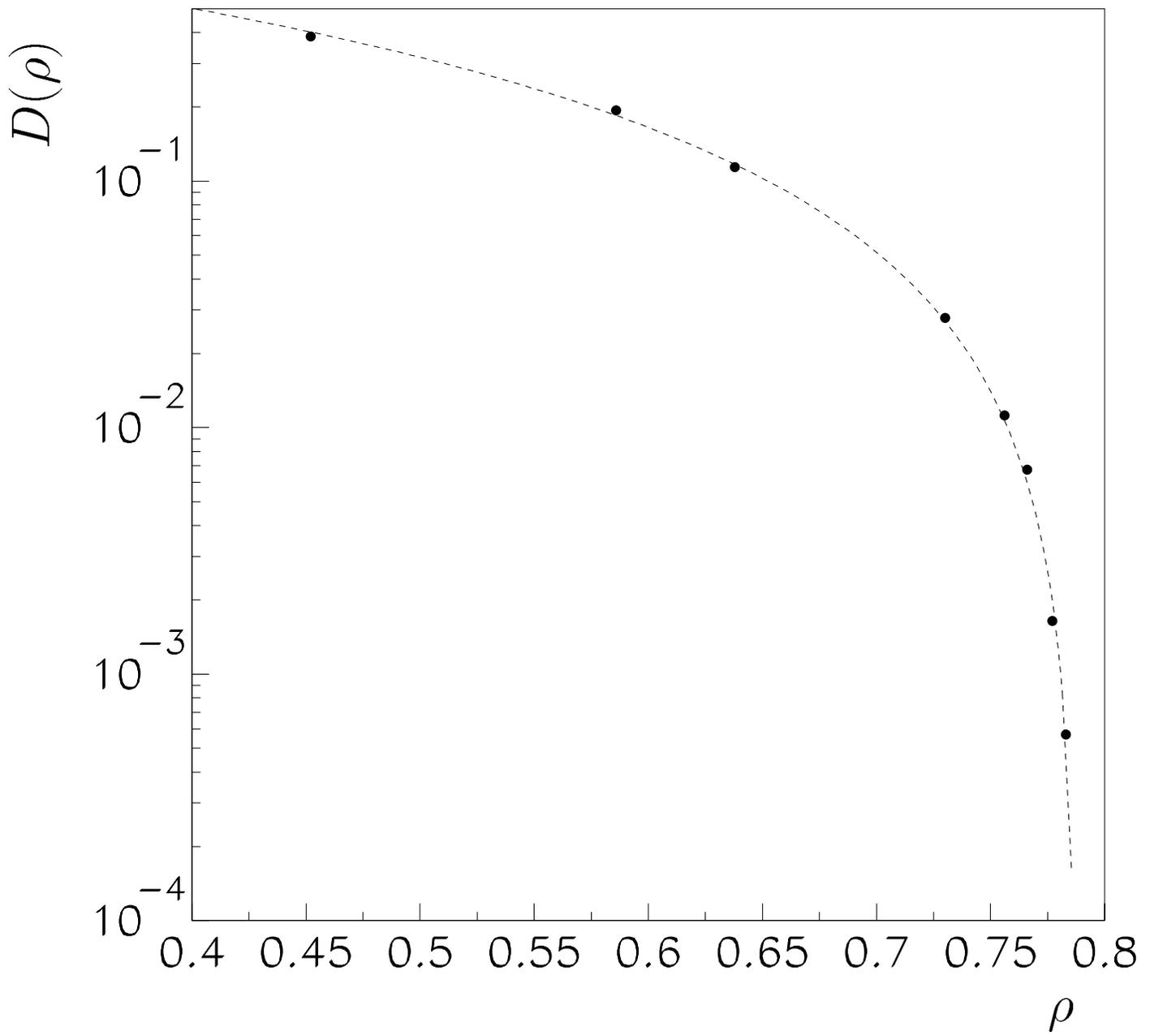
\begin{figure}
\caption{Long time diffusion coefficients $D$ as function of density.
The fit function is $D=2.14\times |\rho-0.7874|^{1.53}$.}
\label{diffusion_fit}
\end{figure}
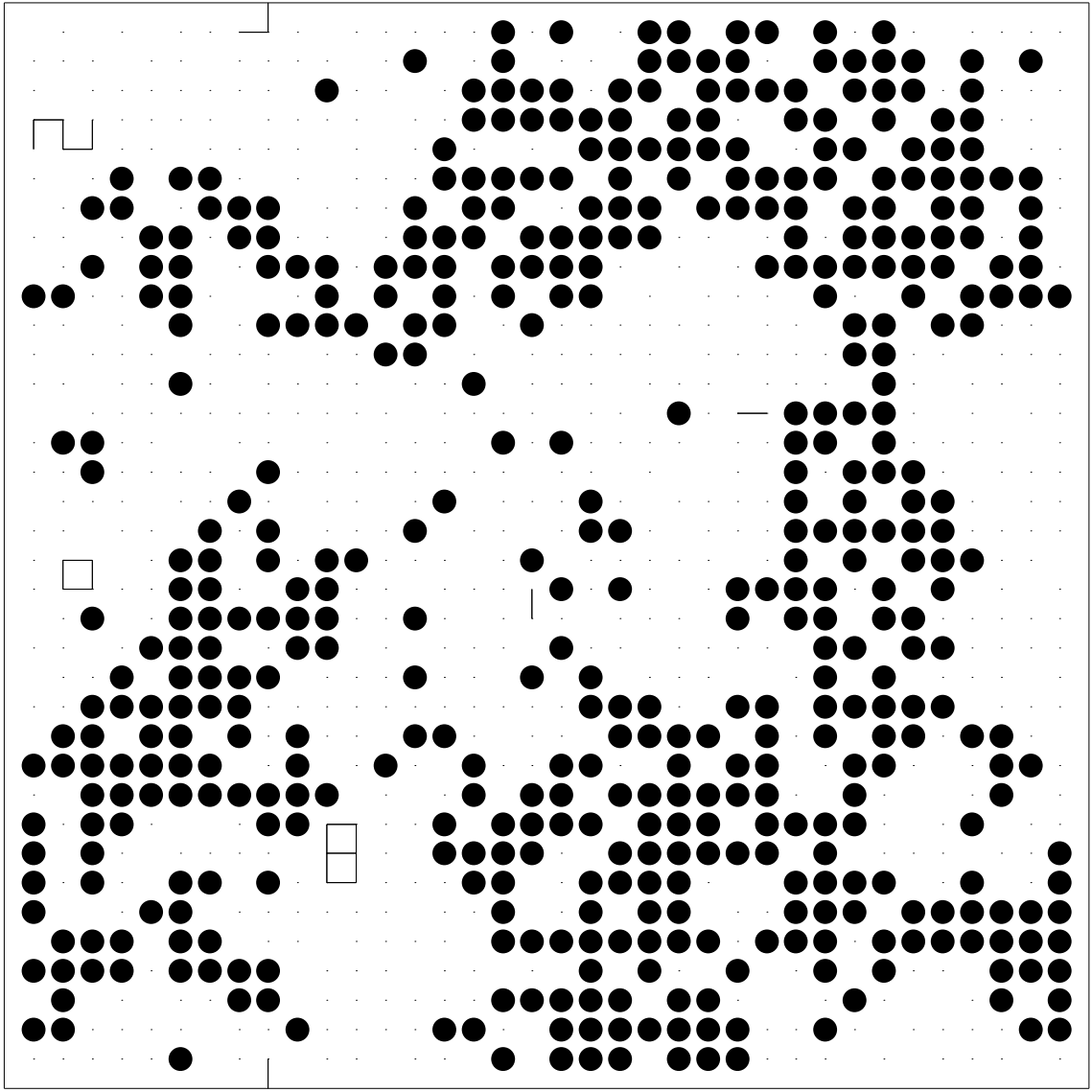
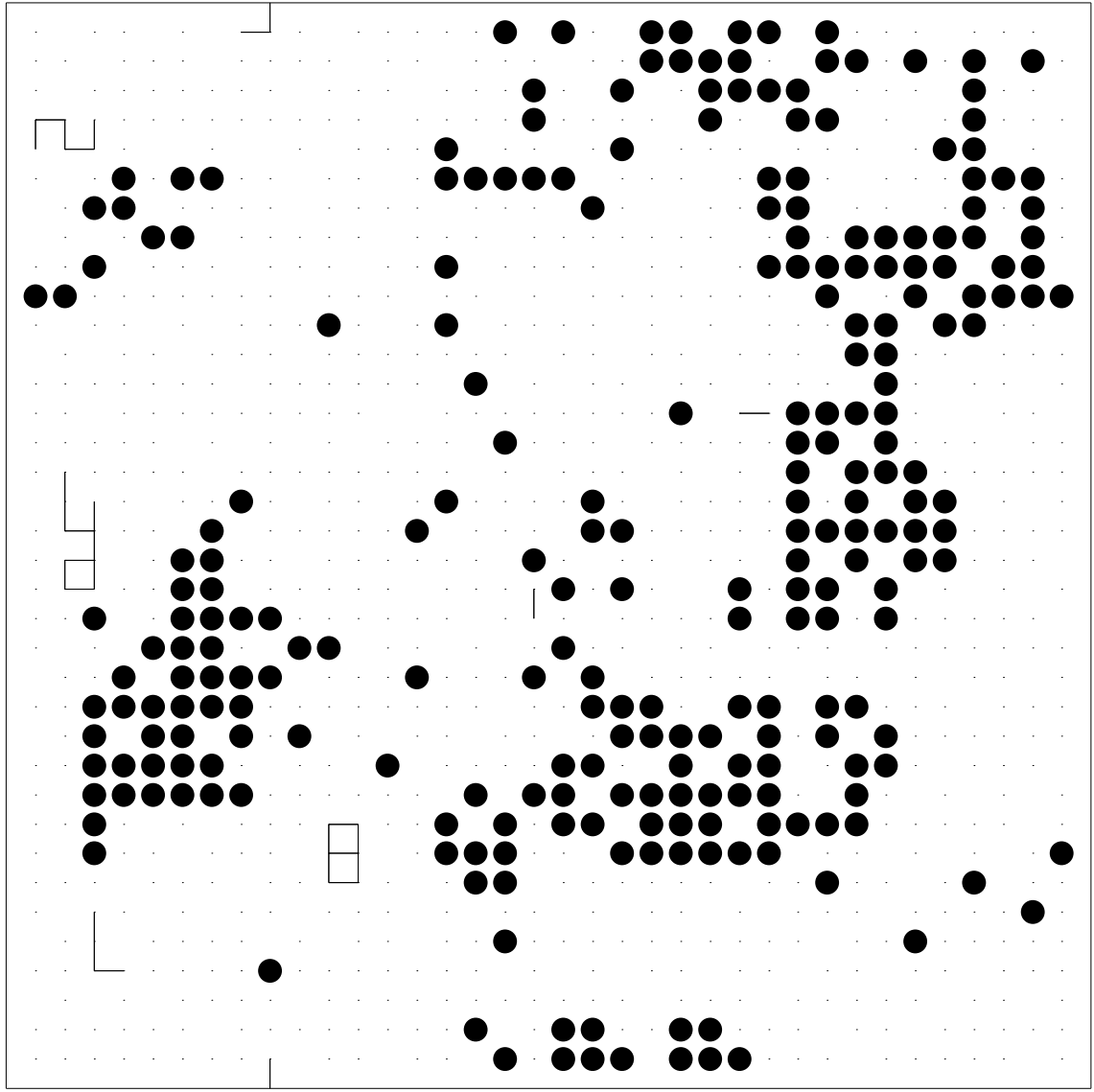
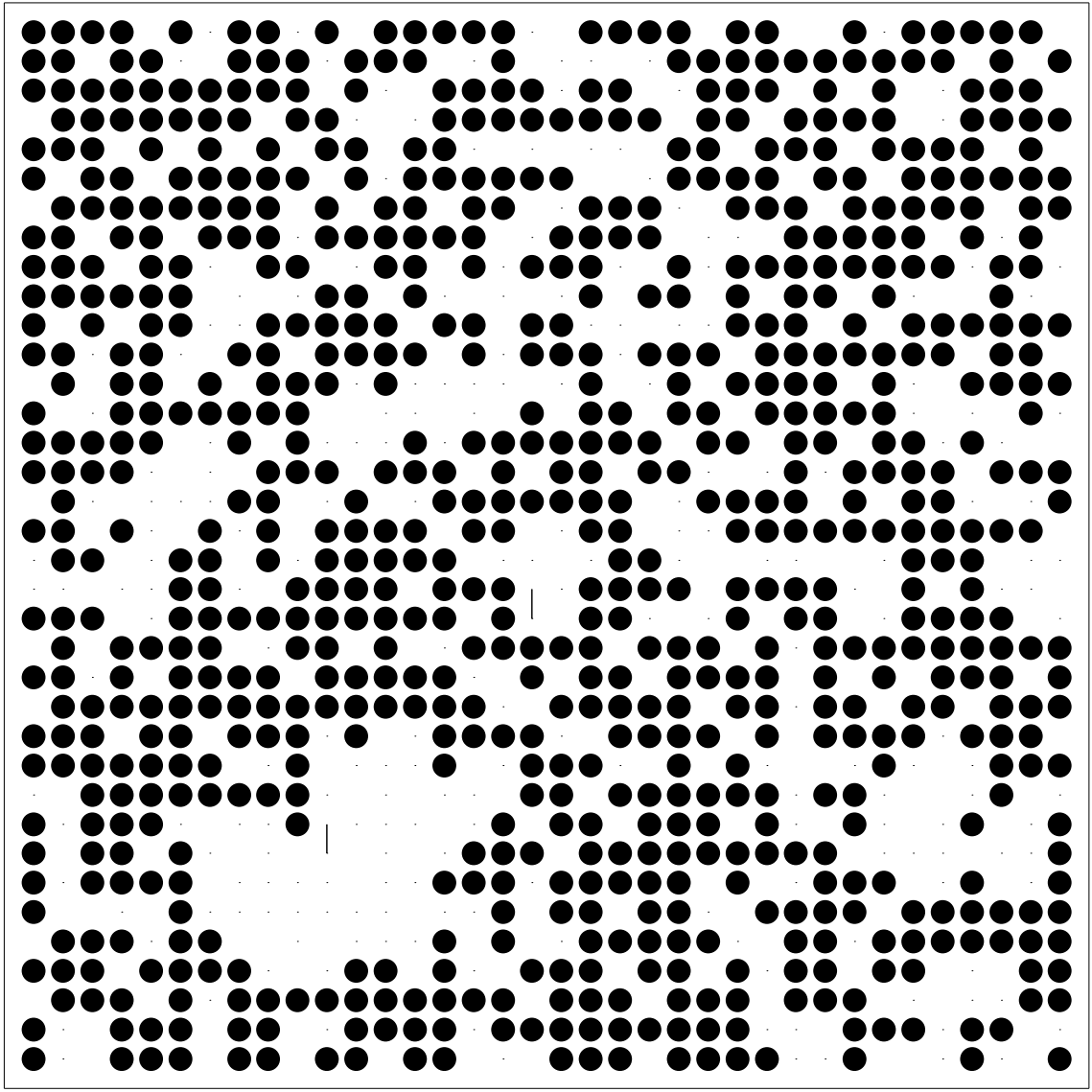
\begin{figure}
\caption{Path of tagged particles
 at density $\rho=0.783$  after $2000$ steps (a), after $4000$ steps (b),
and at $\rho=0.785$ after $2000$ steps (c).
Dots are particles that have moved at least one time in the time
indicated (non frozen domains);
the filled circles represent particles which have never moved in the time
indicated (frozen domains)
The frozen domains
in (c),
are still frozen after $2\times 10^5$ steps. }
\label{diffusion_path}
\end{figure}
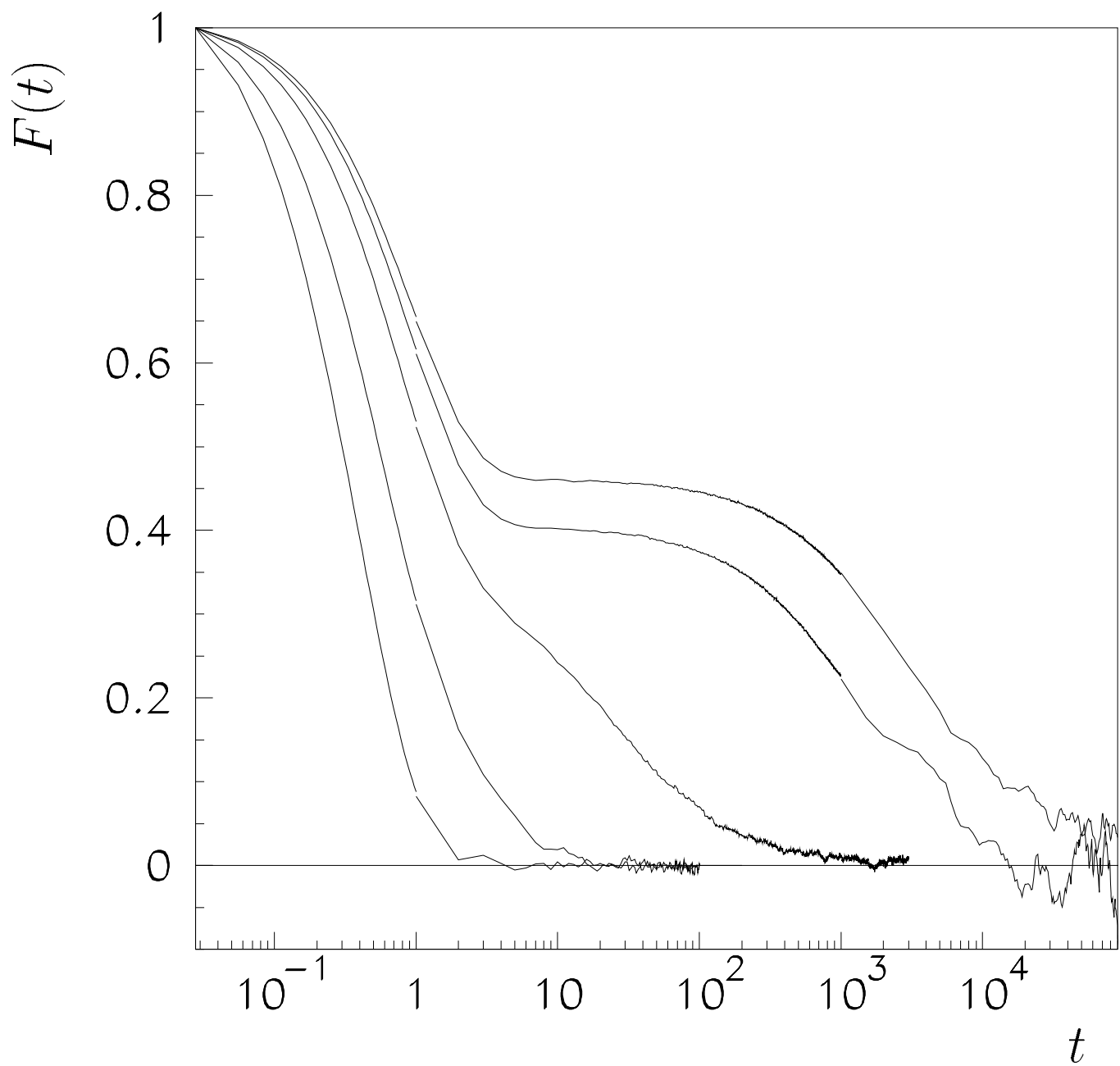
\begin{figure}
\caption{Relaxation functions of the site density,
for the thermalization dynamics, on a $L=36$ square lattice,
for temperatures between $e^{\beta\mu}=1$ and $e^{\beta\mu}=1000$,
corresponding to equilibrium density (from lower to higher relaxation times):
$\rho=0.450$, $0.639$, $0.715$, $0.782$, $0.784$.}
\label{correl_term}
\end{figure}
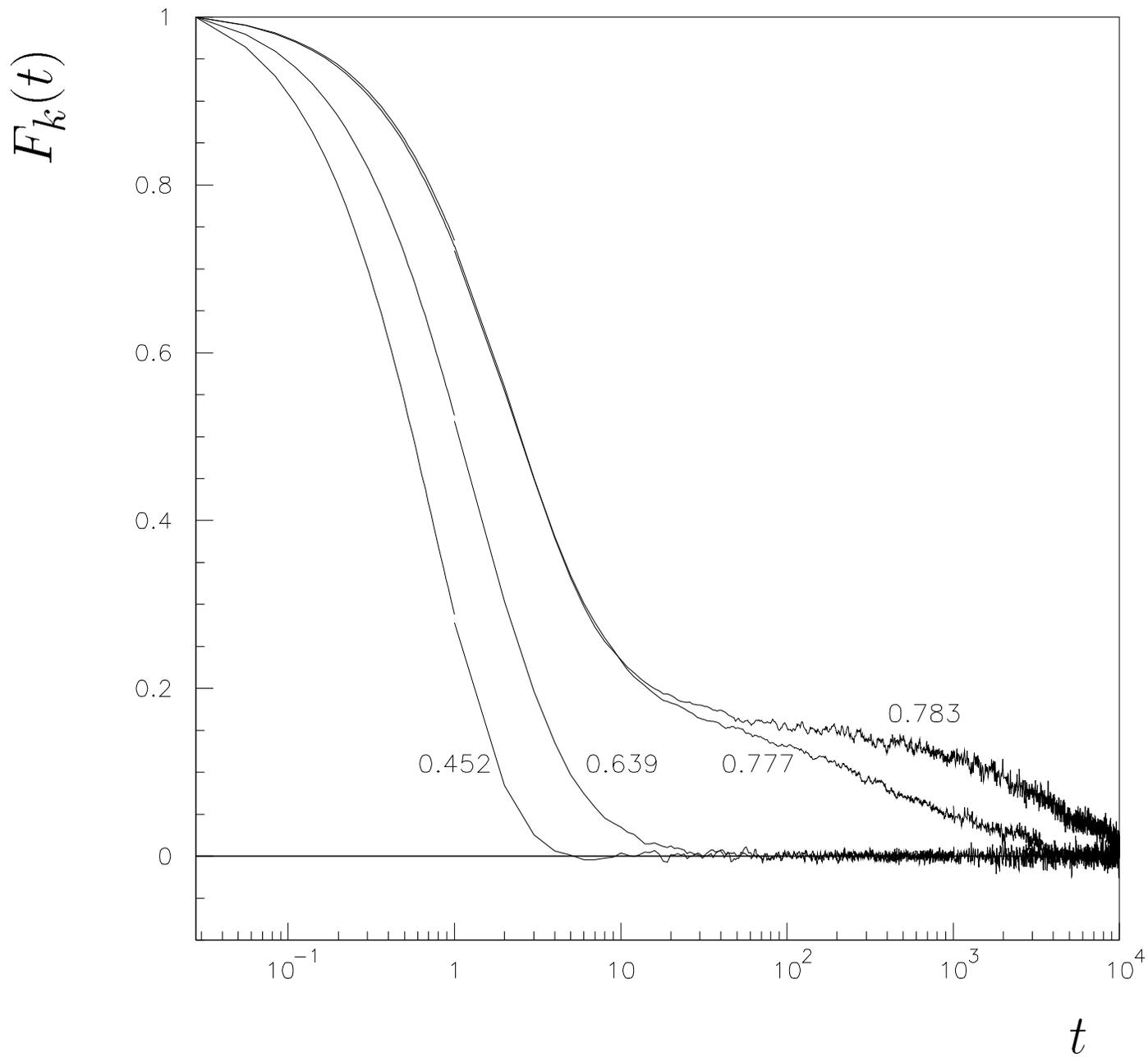
\begin{figure}
\caption{Relaxation functions of the density fluctuations, for the
diffusion dynamics,
for densities $\rho=0.452$, $0.639$, $0.777$, $0.784$,
on a $L=36$ square lattice and $k_x=\pi$, $k_y=0$.}
\label{correl_diff}
\end{figure}
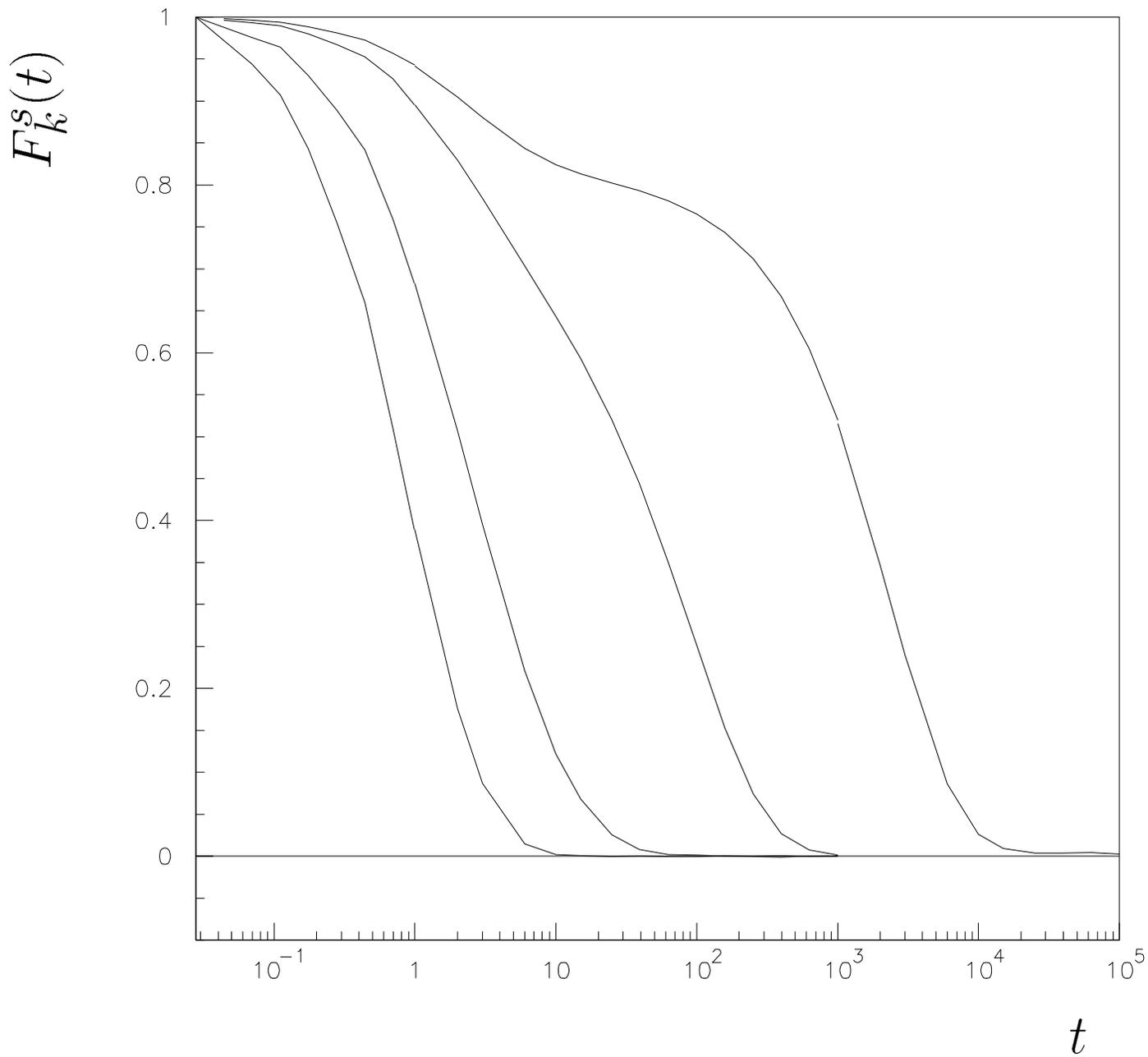
\begin{figure}
\caption{Self part of relaxation functions of the density fluctuations,
for the diffusion dynamics,
for densities $\rho=0.452$, $0.639$, $0.757$, $0.783$ on a
$L=36$ square lattice, with $k_x=\pi$, $k_y=0$.}
\label{correl_self}
\end{figure}
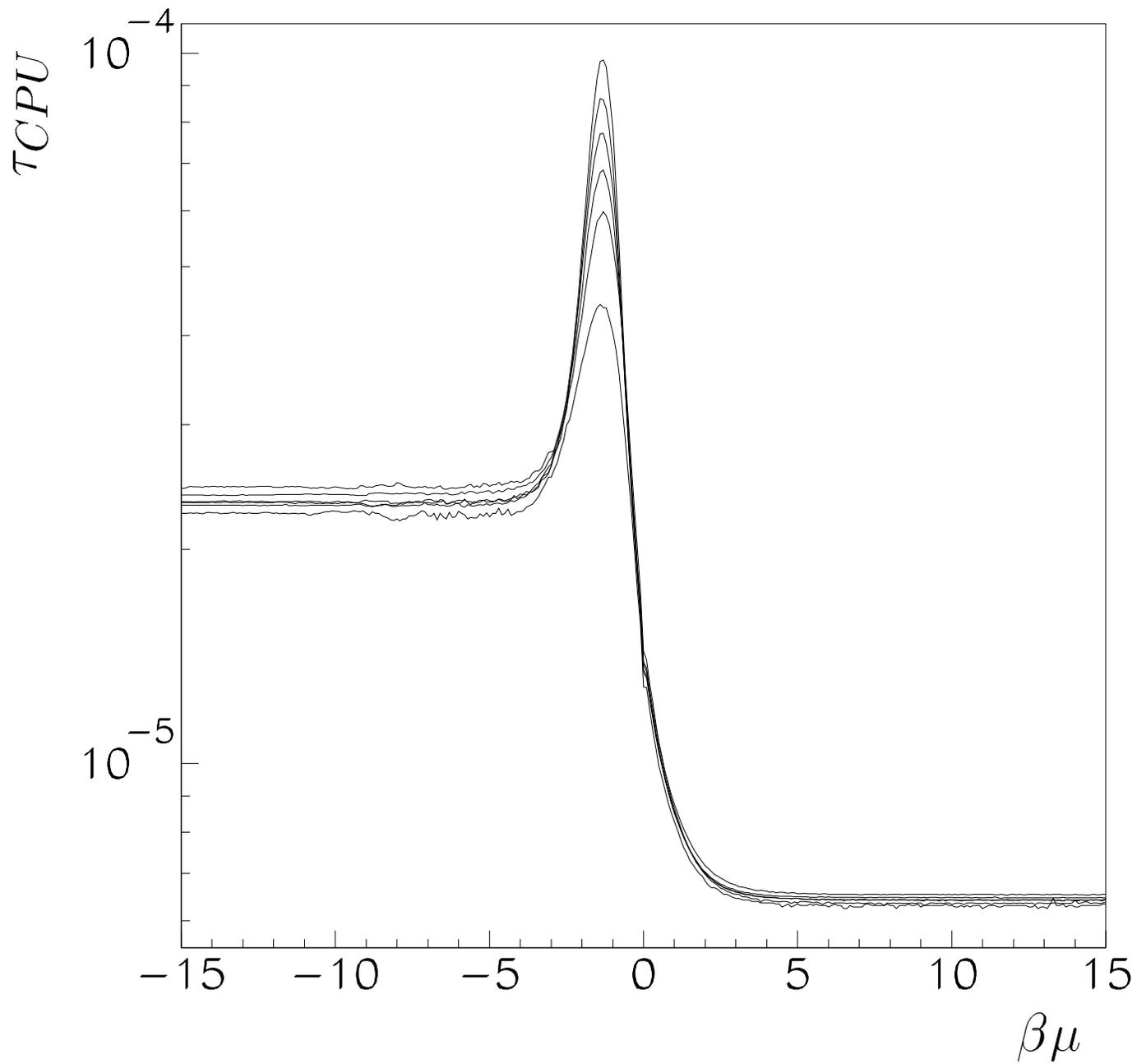
\begin{figure}
\caption{CPU times needed to make a single site update
for lattice sizes $L=24,32,40,48,56,64$.}
\label{CPU_times}
\end{figure}

\begin{references}
\bibitem{foot} If in a given
configuration of bonds there is more than one path of bonds connecting
$i$ and $j$, it is straightforward to show that the resulting phase
$\eta_{ij}$ is path independent since the configuration does not
contain frustrated loops.
\bibitem{nic} It can be shown that this constraint can be induced by the
$\epsilon_{ij}$ when these variables represent interactions between the
particles internal degree of freedom. See M. Nicodemi and A. Coniglio to be
published.
\end{references}
\end{document}